\documentclass[iop]{emulateapj}
\usepackage{apjfonts,natbib,pstricks,pst-grad,amsmath,booktabs,color,epsfig}

\newcommand{\snr}{G350.1$-$0.3}
\newcommand{\cco}{XMMU J172054.5$-$372652}
\newcommand{\chandra}{{\em Chandra}}
\newcommand{\spitzer}{{\em Spitzer}}

\shorttitle{Chandra Observation of G350.1$-$0.3}
\shortauthors{Lovchinsky et al.}
\slugcomment{{\sc Accepted to ApJ:} February 14, 2011}

\begin{document}

\title{A Chandra Observation of Supernova Remnant G350.1$-$0.3 \\
    and Its Central Compact Object}

\author{I.~Lovchinsky\altaffilmark{1}, P.~Slane\altaffilmark{1}, B.~M.~Gaensler\altaffilmark{2},
J.~P.~Hughes\altaffilmark{3}, C.-Y.~Ng\altaffilmark{4},
J.~S.~Lazendic\altaffilmark{5}, J.~D.~Gelfand\altaffilmark{6}, \&
C.~L.~Brogan\altaffilmark{7}}

\altaffiltext{1}{Harvard-Smithsonian Center for Astrophysics, 60 Garden
Street, Cambridge, MA 02138, USA; ilovchin@fas.harvard.edu;
slane@cfa.harvard.edu}
\altaffiltext{2}{Sydney Institute for Astronomy, School of Physics, The 
University of Sydney, NSW 2006, Australia; bryan.gaensler@sydney.edu.au}
\altaffiltext{3}{Rutgers University, The State University of New Jersey, Piscataway, NJ, 
USA; jph@physics.rutgers.edu}
\altaffiltext{4}{Department of Physics, McGill University, Montreal, QC H3A 2T8, Canada; 
ncy@hep.physics.mcgill.ca}
\altaffiltext{5}{School of Physics, Monash University Clayton, VIC 3800, Australia;
Jasmina.Lazendic-Galloway@monash.edu}
\altaffiltext{6}{New York University Abu Dhabi, PO Box 129188, Abu Dhabi, United Arab 
Emirates; jg168@astro.physics.nyu.edu}
\altaffiltext{7}{National Radio Astronomy Observatory, 520 Edgemont Rd, Charlottesville VA, 
22903, USA; cbrogan@cv.nrao.edu}

\begin{abstract}
We present a new \chandra\ observation of supernova remnant (SNR) \snr. The high
resolution X-ray data reveal previously unresolved filamentary structures and
allow us to perform detailed spectroscopy in the diffuse regions of this SNR.
Spectral analysis demonstrates that the region of brightest emission is
dominated by hot, metal-rich ejecta while the ambient material along the
perimeter of the ejecta region and throughout the remnant's western half is
mostly low-temperature, shocked interstellar/circumstellar medium (ISM/CSM) with
solar-type composition. The data reveal that the emission extends far to the
west of the ejecta region and imply a lower limit of 6.6~pc on the diameter of the 
source (at a distance of 4.5~kpc). We show that \snr\ is likely in the free expansion 
(ejecta-dominated) stage and calculate an age of $600 - 1200$~years. The derived
 relationship between the shock velocity and the electron/proton temperature ratio 
 is found to be entirely consistent with that of other SNRs. We perform spectral 
 fits on the X-ray source \cco, a candidate central compact object (CCO), and 
 find that its spectral properties fall within the typical range of other CCOs. 
 We also present archival 24~${\rm\ \mu m}$ data of \snr\ taken with the 
 \emph{Spitzer Space Telescope} during the MIPSGAL galactic survey and find 
 that the infrared and X-ray morphologies are well-correlated. These results 
 help to explain this remnant's peculiar asymmetries and shed new light on its 
 dynamics and evolution.
\end{abstract}


\section{Introduction}

Observations of young supernova remnants (SNRs) provide our best means of relating the thermodynamic properties of shock waves and ejecta distributions to the mechanics of supernovae and their associated progenitors. \chandra\ studies of young remnants have yielded some of the most spectacular advances in SNR research and revealed intriguing and unexpected features of these intricate systems.

\snr\ is a young, luminous SNR in the inner Galaxy. Its nonthermal and linearly polarized radio emission led to its
identification as a SNR by Clark et al. (1973,1975), but a 4.8~GHz Very Large Array radio image (Salter et al. 1986) revealed
a bizarre, asymmetric morphology, unlike that of the typical shell-like structures. In subsequent SNR catalogs, the source
was either removed or listed as a SNR candidate (Green 1991; Whiteoak \& Green 1996). \snr\ has been detected with
\emph{ROSAT} and \emph{ASCA}, where it was labeled 1RXS J172106.9$-$372639 and AX J1721.0$-$3726, respectively (Voges et al.
1999, Sugizaki et al. 2001). An observation with \emph{XMM-Newton} (Gaensler et al. 2008, hereafter G08) revealed the source
to be concentrated in a bright clump, coincident with the region of strongest radio emission. The X-ray spectrum was found to
be well-fit by a two-component model (XSPEC model VPSHOCK + RAYMOND), consisting of a shocked plasma with electron temperature $kT_{e}
\approx 1.5{\rm\ keV}$, ionization timescale $\tau\ \approx 3 \times
10^{11}{\rm\ s\ cm^{-3}}$ and large overabundances of all metals, as well as a collisionally equilibrated, low-temperature ($kT_{e}\approx 0.4{\rm\ keV}$) component with solar
abundances. The best-fit value for interstellar absorption was $N_{H} \approx 3.8 \times 10^{22}{\rm\ cm^{-2}}$. G08 proposed that \snr\ is interacting with a molecular cloud seen along its eastern edge in the $^{12}{\rm CO}$
survey of Bitran et al. (1997). These considerations as well as data from the Southern Galactic Plane Survey
(McClure-Griffiths et al. 2005) were used to derive an approximate distance of 4.5~kpc. Henceforth we scale quantities with
$d_{4.5}$, the distance in units of 4.5 kpc.  

G08 detected an unresolved X-ray source, \cco, to the west of the bright emission and proposed it to be a neutron star, likely a central compact object (CCO), associated with \snr. Its spectrum was found to be a good fit to an absorbed blackbody although a power-law model with an unphysically high photon index ($\Gamma = 5.4$) provided an acceptable fit as well. G08 carried out a search for possible pulsations but found no periodic signals in the range of 146~ms to 1~hr.

In this paper, we present a detailed study of \snr\ and its candidate CCO. In \S\ 2 we summarize the \chandra\ and \spitzer\ observations and data reduction. \S\ 3.1 features a discussion of the X-ray and infrared imaging and a comparison of various spectral fits. In \S\ 3.2 we use the parameters derived from the models to analyze the dynamics and evolution of \snr. A discussion of the compact source \cco\ follows in \S\ 3.3. In \S\ 4 we interpret the results of these investigations and attempt to present a coherent and self-consistent picture of \snr. We conclude with \S\ 5, a summary of our findings.

\begin{figure}[t]
\epsscale{1.18}
\plotone{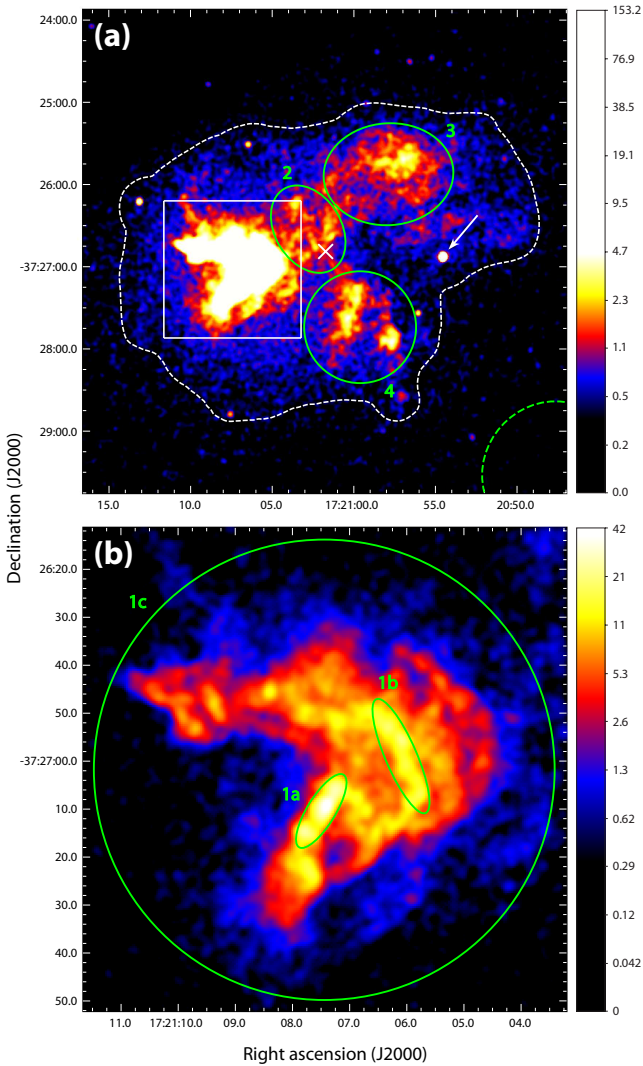}
\caption{
Panel \emph{a} shows a \chandra\ image of \snr\ in the $0.5 - 8.0$~keV band. 
The image has been smoothed by a Gaussian kernel with a $\sigma$ of 3 pixels (1.48~arcsec). 
The detailed structure of the bright emission region in the east is not visible due to the 
high contrast. The dashed contour surrounding the SNR is at a level of 3$\sigma$ above the 
background and shows the apparent extent of the emission. The white cross indicates the centroid 
of the contour and is regarded as the putative center of the SNR throughout this analysis.
Regions $2 - 4$ were used for spectral extraction (as shown in Figure 3) with parameters
listed in Table 1. The dashed circle, partly shown in the southwest corner, was used to 
subtract the local background. The arrow indicates the position of the candidate CCO \cco.
Panel \emph{b} shows an enlarged image of the bright eastern emission indicated by 
the square region in panel \emph{a}. Regions 1a, 1b and 1c were used for spectral extraction
(as shown in Figure 2), with parameters for regions 1a and 1b listed in Table 1.
The colorbars (in units of counts/pixel) indicate brightness levels. 
}
\end{figure}
\section{Observations}

In order to obtain a high-resolution map of the SNR emission structure, study
its spectral properties in unprecedented detail and obtain tighter constraints on the spectral fit parameters,
we observed \snr\ on 2009 May 21 and 22 for 84~ks with the 
Advanced CCD Imaging Spectrometer (ACIS-S) on board the \emph{Chandra X-ray Observatory} (observation ID 10102). 
Standard data reduction was performed to remove hot pixels and flares, resulting in an 
effective exposure time of 83~ks. The 24~${\rm\ \mu m}$ and 70~${\rm\ \mu m}$
images of the spatial region corresponding to \snr\ were obtained during MIPSGAL,
an infrared survey of the Galactic plane (Carey
et al. 2009). The target was observed (in both bands) on 2006 October 06 with the
Multiband Infrared Photometer for Spitzer (MIPS) instrument on board the \emph{Spitzer
Space Telescope}. The raw data were calibrated using the
standard BCD (basic calibrated data) pipeline to correct for bright pixels
and flares resulting in exposure times of 2.6~s (24~${\rm\ \mu m}$) and
3.15~s (70~${\rm\ \mu m}$). The 70~${\rm\ \mu m}$ data were additionally filtered
to mitigate instrumental effects (stimflash latents and residual background
drifts). An 8~${\rm\ \mu m}$ image was obtained during GLIMPSE, an infrared survey of the inner Milky Way
Galaxy (Churchwell et al. 2009).  The target was observed on 2004 September 07 with the InfraRed Array Camera 
(IRAC) on board the \emph{Spitzer
Space Telescope}. The raw data were calibrated using the
standard BCD pipeline to correct for hot pixels
and flares resulting in an exposure time of 1.2~s.

\section{Analysis}
\subsection{Imaging and Spectroscopy}

A \chandra\ image of \snr\ is shown in Figure 1a. We filtered the data on the
0.5 - 8.0~keV energy band and smoothed the image by applying a Gaussian kernel
with a $\sigma$ of 3 pixels (1.48 arcsec). The source is dominated by an
extended bright clump of material in the east with fainter emission extending
far to the west. The dashed contour, shown in the figure, surrounds the SNR at a level of 
3$\sigma$ above the background and indicates the apparent extent of the emission. Throughout
our analysis, we adopt a source radius of $R = 3.3 d_{4.5}$~pc (2.5~arcmin) and
assume, following the argument of G08, that $d_{4.5} = 1$. Although the data 
reveal an elliptical shell-like structure, a large region in the southwest is markedly
fainter than the adjacent emission with a flux only a factor of $\sim 1.5$ above the background.
The unresolved X-ray source \cco\ (indicated by the arrow in Figure 1a) is seen far to the west of the bright
emission, significantly displaced from the apparent center of the SNR (marked by
the white cross in Figure 1a).  Figure 1b shows the detailed structure of the ejecta 
region and reveals previously unresolved filamentary structures
running across its center. We find that the six regions enumerated in
Figures 1a and 1b represent the range of spectral characteristics found in this
remnant. The spectra and the corresponding response files were produced using
the \emph{specextract} script in Ciao 4.2. We subtracted the local background
spectrum from a source-free region of the detector, partly shown as the dashed
region in the southwest corner of Figure 1a, and grouped the data to a minimum
of 15 counts per bin. We fit each spectrum by an absorbed nonequilibrium
ionization (NEI) plane-parallel shock model with variable abundances (XSPEC
model "VNEI"; Borkowski et al. 2001). Elements below Mg and above Fe were fixed
at solar abundances since their contributions in the fitted bandpass are
negligible. The uncertainties were calculated at 90\% confidence ($1.6 \sigma$).
Although fixing the absorption at an average global value of $\sim 4.0 \times
10^{22}{\rm\ cm^{-2}}$ produced acceptable fits for regions 1a, 1b, 2, 3 and 4
(the exception is region 1c and is discussed below), we allow $N_{H}$ to vary in 
order to account for possible small-scale variations in the
absorbing column. The spectra and fitted models for regions 1a, 1b and 1c are shown in Figure 2 with
parameters (for regions 1a and 1b) listed in Table 1. The fits for regions 2, 3
and 4 are shown in Figure 3, with parameters listed in Table 1.

\begin{figure}[t]
\epsscale{1.18}
\plotone{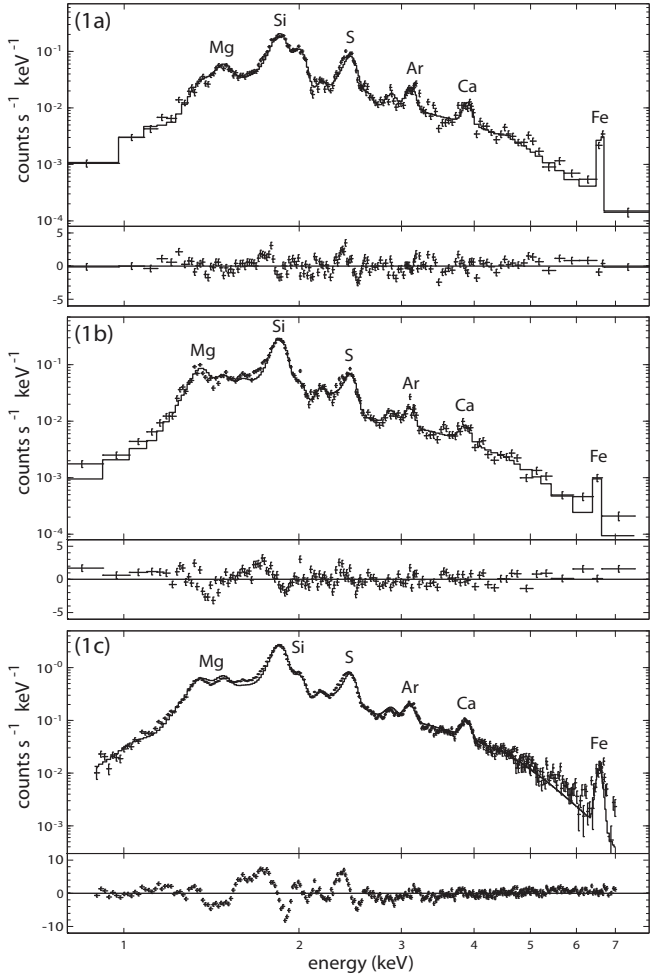}
\caption{
Plots of spectra extracted from regions 1a, 1b and 1c overlaid with the fitted VNEI models and residuals. 
The NEI model in XSPEC does not include atomic data for the argon emission; a Gaussian component was added 
to the model to account for this feature in the spectral fitting. The fit parameters for regions 1a and 1b are
listed in Table 1. 
}
\end{figure}

\begin{figure}[t]
\epsscale{1.18}
\plotone{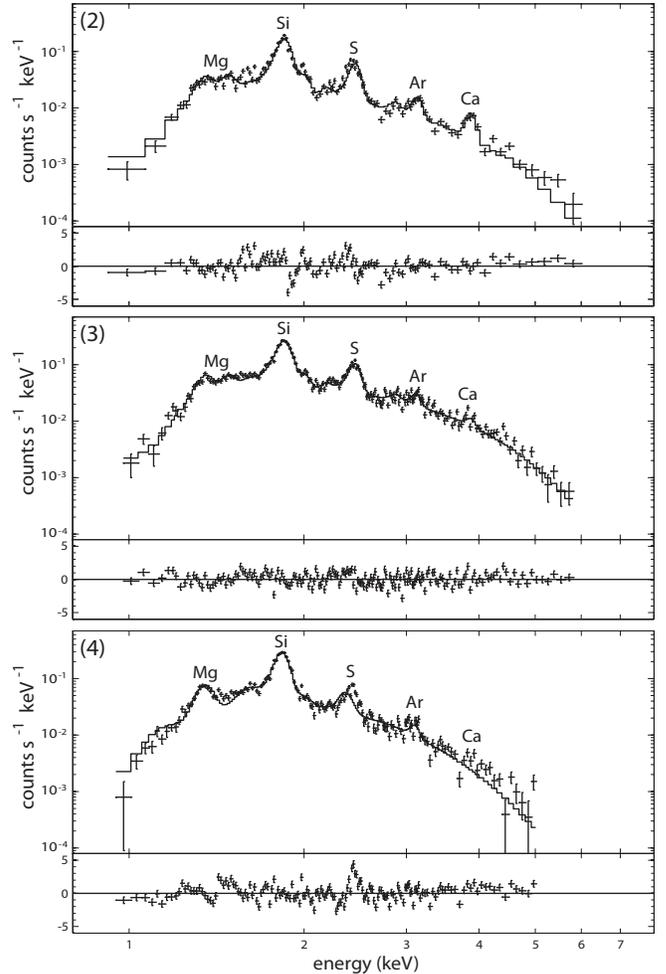}
\caption{
Plots of spectra extracted from regions 2, 3 and 4 overlaid with the fitted VNEI models and residuals. The fit parameters are
listed in Table 1.
}
\end{figure}

The spectrum of region 1a (extracted from one of the bright filaments in Figure
1b) is roughly representative of that found throughout the bright
eastern emission. It can be fit by an absorbed VNEI model with column density
$N_{H} \approx 3.8 \times 10^{22}{\rm\ cm^{-2}}$, electron temperature $kT_{e}
\approx 1.4{\rm\ keV}$, ionization timescale $\tau\ \approx 2.2 \times
10^{11}{\rm\ s\ cm^{-3}}$ and large overabundances of Mg, Si, S, Ca and Fe.
These characteristics unambiguously demonstrate the presence of hot, metal-rich
ejecta. The spectrum of region 1b (extracted from the other bright filament) is similar to that of region 1a although the Mg and
Si lines clearly indicate a somewhat lower ionization timescale.  Although morphologically well-defined, 
we find that the spectral characteristics of the filaments are basically consistent with the surrounding regions. 

\begin{table*}
\begin{center}
\caption{\footnotesize{VNEI parameters for spectral fits of regions 1a, 1b, 2, 3, 4}}
\label{tab:dens}
\begin{tabular}{lrrrrrr}
\toprule
\toprule

Parameter &&Region 1a&Region 1b&Region 2&Region 3&Region 4  \\

\midrule
$N_{\rm H}$\tablenotemark{a} ($10^{22}$\,cm$^{-2}$) && $3.8_{-0.2}^{+0.4}$ & $4.0_{-0.1}^{+1.3}$ & $3.9_{-0.2}^{+0.3}$ & $4.4_{-0.2}^{+0.3}$ & $4.3_{-0.2}^{+0.2}$ \\
\noalign{\smallskip}
$kT_{e}$ (keV) && $1.4_{-0.2}^{+0.2}$ & $1.2_{-0.1}^{+0.2}$ & $0.8_{-0.1}^{+0.1}$
& $0.9_{-0.1}^{+0.1}$ & $0.54_{-0.04}^{+0.04}$ \\
\noalign{\smallskip}
Mg && $5.8_{-1.8}^{+8.2}$ & $8.6_{-1.8}^{+1.9}$ & $2.6_{-0.7}^{+1.1}$ & $2.2_{-0.6}^{+0.9}$ & $1.1_{-0.2}^{+0.2}$ \\
\noalign{\smallskip}
Si && $8.6_{-1.8}^{+9.3}$ & $7.2_{-1.4}^{+1.4}$ & $2.8_{-0.6}^{+0.5}$ & $1.6_{-0.2}^{+0.3}$ & $1.7_{-0.1}^{+0.3}$ \\
\noalign{\smallskip}
S && $4.4_{-0.9}^{+4.4}$ & $3.0_{-0.6}^{+0.6}$ & $2.3_{-0.4}^{+0.4}$ & $1.2_{-0.2}^{+0.2}$ & $0.5_{-0.1}^{+0.2}$ \\
\noalign{\smallskip}
Ca && $6.1_{-1.8}^{+2.0}$ & $4.9_{-2.3}^{+2.3}$ & $6.3_{-2.4}^{+3.1}$ & $1.1_{-0.9}^{+1.0}$ & $0_{--}^{+4.9}$ \\
\noalign{\smallskip}
Fe\tablenotemark{b} && $4.3_{-1.6}^{+8.1}$ & $4.3_{-1.4}^{+1.4}$ & (1) & (1) & (1) \\
\noalign{\smallskip}
$\tau$ ($10^{11}{\rm\ s\ cm^{-3}}$) && $2.2_{-0.3}^{+0.6}$ & $0.92_{-0.01}^{+0.10}$ & $4.5_{-1.2}^{+2.4}$ & $2.7_{-0.5}^{+0.5}$ & $0.13_{-0.02}^{+0.04}$ \\
\noalign{\smallskip}
Flux\tablenotemark{c} ($10^{-13}{\rm\ ergs\ cm^{-2}\ s^{-1}}$) && 8.9 & 8.2 & 5.4 & 11 & 7.2 \\
\noalign{\smallskip}
$\chi_{\nu}^{2}$/dof && 1.94/157 & 2.16/144 & 1.89/115 & 1.55/178 & 2.27/154 \\
\bottomrule
\noalign{\smallskip}
\multicolumn{7}{p{1.22\columnwidth}}{\footnotesize{{\bf Note.} --- All uncertainties are statistical errors at 90\% confidence, 1.6$\sigma$. Elements below Mg and above Fe were fixed at solar because their contributions in the fitted bandpass are small}}\\
\multicolumn{7}{p{1.22\columnwidth}}{\footnotesize{$^{\text{a}}$The absorption was calculated using the model of Wilms et al. (2000)}}\\
\multicolumn{7}{p{1.22\columnwidth}}{\footnotesize{$^{\text{b}}$(1) indicates that the elemental abundance was fixed at solar}}\\
\multicolumn{7}{p{1.22\columnwidth}}{\footnotesize{$^{\text{c}}$The values listed are the unabsorbed fluxes over the energy range $0.5 - 8.0$~keV}}\\
\end{tabular}
\end{center}
\end{table*}

Although G08 found evidence for variations in elemental abundances between nearby regions 
within the bright eastern emission, our
spectral fits of small regions in this vicinity show no significant variations
within uncertainties. In order to compare our results with those of the
\emph{XMM-Newton} observation, we attempted to fit the entire ejecta-dominated emission
(region 1c: the same area as fitted by G08) using a VNEI model. As can be seen from the high residuals, this model does not provide
a good fit, and hence, we do not display the fit parameters. However, the best-fit parameters are roughly consistent with those of regions 1a and 1b, indicating that the extremely large number of
X-ray photon counts ($\sim 100,000$) and small error bars likely magnify small deficiencies in the model and result in the poor $\chi^{2}$
statistics. In
addition, it is possible that small variations in temperature and elemental abundances exist, but are not discernible with the
statistics in our spectra from small-scale, spatially-resolved regions.
Using the two-component model VPSHOCK + RAYMOND (the same as that used by G08 - their fit results are summarized in \S\ 1) for region 1c also fails to provide
a good fit. The inconsistencies between our results and those of G08 are likely explained simply by the fact that the long
exposure in the \chandra\ observation exposes weaknesses in the models that were previously hidden due to 
the fact that the \emph{XMM-Newton} spectrum has fewer photon counts.

Region 3 is largely representative of the spectral features found along the perimeter of the ejecta region and throughout the
remnant's western half. The spectrum can be described by a VNEI model with absorbing column $N_{H} \approx 4.4 \times
10^{22}{\rm\ cm^{-2}}$, electron temperature $kT_{e} \approx 0.85{\rm\ keV}$, ionization timescale $\tau\ \approx 2.7 \times
10^{11}{\rm\ s\ cm^{-3}}$ and near-solar abundances of all metals. These results suggest an interstellar/circumstellar
origin. Although otherwise similar, the fit for region 4 indicates a somewhat lower temperature and an ionization timescale
of $\tau\ \approx 1.3 \times 10^{10}{\rm\ s\ cm^{-3}}$, lower than that of region 3 by a factor of 20. The centroid of the Si line in the
region 4 spectrum is at a slightly lower energy than that of region 3 and since the goodness of fit statistics are dominated by the Si
line (the error bars are smallest in this region), it is this feature that causes the drastic difference in the ionization timescale. 
Region 2, adjacent to the bright eastern emission,
has somewhat enhanced abundances of Mg, Si and S and a highly enhanced abundance of Ca. Although similar in temperature to
region 3, the overabundance of metals suggests the possible presence of ejecta. 

The model histogram for the spectrum of region 4 slightly misses the centroid of
the S line complex, suggesting that the several bright patches within this region
likely vary in their ionization states. Fitting the individual clumps, however, leads to a similar result with the added
handicap of poor statistics in the spectra. We thus present the fit to the composite region with the understanding that a
deeper exposure is necessary to characterize the spectrum more precisely. The spectral fits to regions 1b, 2 and 4 do not 
formally correspond to good fits based on the $\chi^2$ statistics. In the case of region 1b, this deficiency is due to 
the fact that the model histogram misses the minimum of the Mg line and is too low on the low-energy shoulder of the Si line. In the
region 2 spectrum, the fit is too low on the maximum of the Si line, as can be seen from the high residuals in this area.
We acknowledge these weaknesses in the models and use them to simply make qualitative statements about the approximate distribution
of elemental abundances, temperatures and ionization states, rather than using them to derive precise numerical results.

A 24~${\rm\ \mu m}$ infrared image of \snr, overlaid with the contours from the
X-ray data, is shown in Figure 4a. The two
bright clumps in the east and the south correspond to regions 1c and 4 respectively. Most notably,
the upper filament from the X-ray image (region 1b) appears to outline precisely the eastern edge of the infrared emission while the
lower filament (region 1a) is notably absent at 24~${\rm\ \mu m}$. An enlarged 24 um image of the ejecta-
dominated region, overlaid with the X-ray contours, is shown in Figure 4b. 
While the angular 
resolution at 24 um is much coarser than that provided by \chandra, these 
images are sufficiently sensitive to identify two distinct regions of 
emission with clear morphological similarities to the X-ray emission from 
the SNR (regions 1c and 4 in Figure 1). Figures 4c and 4d show IRAC 8~${\rm\ \mu m}$ and MIPS 70~${\rm\ \mu m}$ images 
of the spatial region corresponding to \snr. We do not detect any significant
emission from \snr\ at 8~${\rm\ \mu m}$ and although we see 70~${\rm\ \mu m}$ emission in regions
corresponding to the SNR, the emission does not differ significantly from
regions in the 
surrounding medium.  With the current resolution we cannot determine 
whether or not some of the 70~${\rm\ \mu m}$ emission is associated with the SNR. We searched
the 3.6~${\rm\ \mu m}$, 4.5~${\rm\ \mu m}$ and 5.8~${\rm\
\mu m}$ IRAC bands but found no apparent emission from \snr\ or \cco. 

\begin{figure*}[t]
\epsscale{1.16}
\plottwo{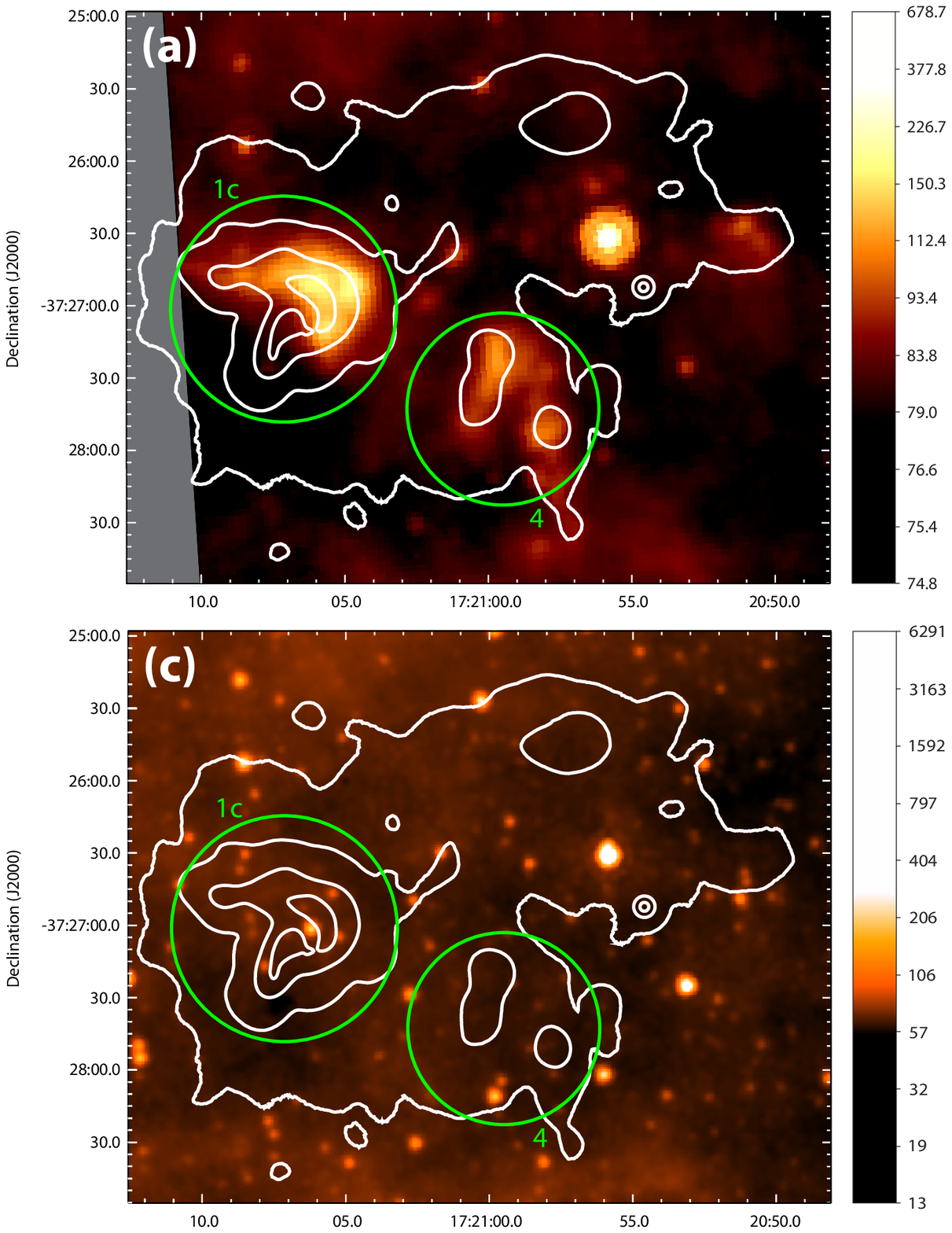}{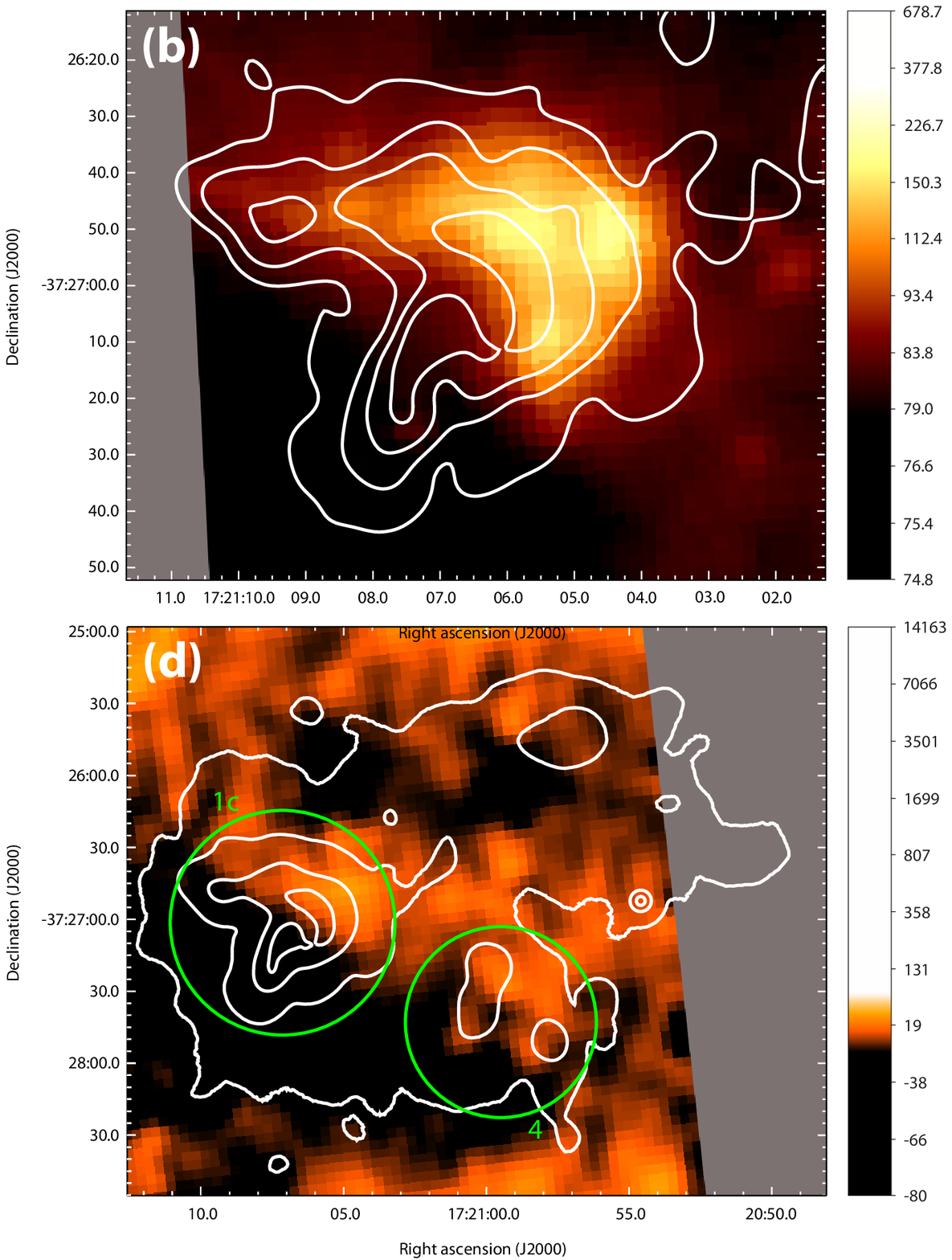}
\caption{
Panel \emph{a} shows a 24~${\rm\ \mu m}$ image of \snr\ taken during the 
MIPSGAL Galactic survey. The integrated flux of regions 1c and 4 is 
$\sim2.5$~Jy and $\sim0.5$~Jy 
respectively. Panel \emph{b} shows an enlarged image of the region corresponding
to the ejecta-dominated emission with corresponding X-ray contours at levels of
1.1, 2.5, 7.0 and 13 counts/pixel. Panels \emph{c} and \emph{d} show 8~${\rm\ \mu m}$ 
and 70~${\rm\ \mu m}$ images of the same spatial region as panel \emph{a}. The
contours in panels \emph{a}, \emph{c} and \emph{d} are at levels of 0.55, 1.1,
5.0 and 13 counts/pixels and correspond to the X-ray data from Figure 1a.
Regions 1c and 4 correspond to the same regions as in Figure 1. The colorbars 
indicate brightness levels in units of MJy/sr. 
}
\end{figure*}

\subsection{Dynamics and Evolution}

We analyze the dynamics and evolution of \snr\ using the numerical study of nonradiative SNRs by Truelove and McKee (1999),
hereafter TM99. We fit the X-ray spectrum of the SNR's entire western half by an absorbed VNEI model in order to derive the
average post-shock electron density and the mass of the swept-up material. The parameters of the model are roughly
intermediate to those of regions 3 and 4. We assume a spherical half-shell with thickness $R/12$ for the geometry and a value
of 4 for the compression ratio of post-shock and pre-shock densities. We also take the ratio between the electron and atomic
hydrogen densities to be ${n_{e} / n_{H} = 1.2}$, valid for cosmic abundances.
With the normalization parameter from the fit ($\sim 0.056$), we use the relation 
\begin{equation}
\frac{n_e n_H V}{4 \pi d^2} = 0.056\times 10^{14} {\rm\ cm}^{-5},
\end{equation}
where $V$ is the estimated volume and $d$ is the distance, to derive an
approximate post-shock electron density $n_{e} \approx 5.6 d_{4.5}^{-1/2}{\rm\ cm^{-3}}$ and a
corresponding swept-up mass of $\sim 2 d_{4.5}^{-3/2}{\rm\ M_{\odot}}$, with the obvious caveat that the uncertainties are
unquantifiable since the visible emission --- and hence the observed radius --- is only a lower constraint on the true size
of the SNR. G08 derived a substantially higher density through two independent methods. However, the first calculation
assumes ionization equilibrium, inferred from using a two-component model (which we were unable to fit, as discussed in \S\
3.1). The second calculation assumes a Sedov solution --- an assumption that is likely not valid for reasons discussed below.
If the molecular cloud interaction scenario (as proposed by G08) is correct, 
the density throughout the ejecta-dominated half of the SNR is likely to be 
higher than that in the western half. Hence, extrapolating our estimate to the
eastern half of the SNR (for a total of $\sim 4 d_{4.5}^{-3/2}{\rm\ M_{\odot}}$) is likely an 
underestimate of the true swept-up mass, although this figure can be used
as a rough lower bound. 

For a progenitor with an ejecta density profile $\varpropto r^{-7}$ (typical of
core-collapse supernovae), TM99 demonstrate that the radius and age at which a
free-expanding SNR enters the Sedov-Taylor (ST) phase are given by
\begin{equation}
R_{ST} = 0.881 M_{ej}^{1/3} \rho_{0}^{-1/3}
\end{equation}
and
\begin{equation}
t_{ST} = 0.732 E_{SN}^{-1/2} M_{ej}^{5/6} \rho_{0}^{-1/3},
\end{equation}
where $M_{ej}$ is the ejecta mass and $E_{SN}$ is the explosion energy. Here $\rho_{0}$ is the pre-shock mass density and is given
by
\begin{equation}
\rho_{0} = {\displaystyle \frac{1.4}{(1.2)(4)}} m_{H} n_{e} = 0.30 m_{H} n_{e},
\end{equation}
where $m_{H}$ is the mass of the hydrogen atom, 1.4 is the equivalent molecular
weight of the hydrogen and helium mixture, 
(assuming cosmic abundances) and the factors 1.2 and 4 are the electron/hydrogen density
ratio and post-shock/pre-shock compression ratio, respectively. Using typical values of $M_{ej} = 4 - 14{\rm\ M_{\odot}}$ for
the ejecta mass consistent with the formation of a neutron star (Woosley et al. 2002), we calculate $R_{ST}$ to be $4.1 -
6.2$~pc, somewhat larger than the observed radius for \snr\ ($R = 3.3 d_{4.5}$~pc). Assuming that the observed radius of the SNR is approximately equal to its
true radius (in \S\ 4 we discuss the scenario in which this 
assumption is not valid), this result suggests that \snr\ is likely in the free expansion
(ejecta-dominated) stage and has not yet entered the ST phase. 

\begin{figure}[t]
\epsscale{1.15}
\plotone{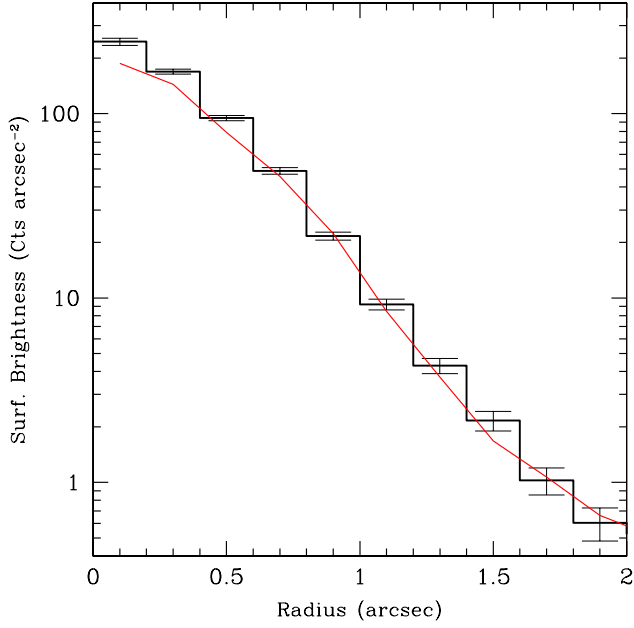}
\caption{
Azimuthally-averaged radial profile of CCO \cco\ in the $0.5 - 8.0$~keV energy range. The histogram with the vertical error
bars represents the data while the curve is the model point spread function (PSF).
}
\end{figure}

If we make an assumption about the explosion energy, we can determine the age using the relation
\begin{equation}
t = 0.988 t_{ST} \left({\displaystyle \frac{R}{R_{ST}}}\right)^{7/4},
\end{equation}
as derived from TM99. We explore the parameter space for $M_{ej} = 4 - 14{\rm\ M_{\odot}}$ and $E_{SN} = 0.5 - 1.0 \times 10^{51}$~ergs and derive an age of $t = 600 - 1200$~years. The shock velocity and proton temperature follow from\footnote{This expression for the shock velocity comes from TM99 Table 7. It should be noted, however, that the authors overlooked the minus sign in the exponent.}
\begin{equation}
v_{b} = {\displaystyle \frac{dR}{dt}} = 0.576 \left({\displaystyle \frac{R_{ST}}{t_{ST}}}\right)\left({\displaystyle \frac{t}{t_{ST}}}\right)^{-3/7}
\end{equation}
and
\begin{equation}
kT_{p} = 0.11 m_{H} v_{b}^{2} .
\end{equation}
The results of these calculations are summarized in Table 2. Using $kT_{e} \approx 0.7$~keV for the electron temperature of the ISM component, the average value from our fits, the inferred electron-to-proton temperature ratio is $0.07 - 0.27$. This is in excellent agreement with measured values of this ratio for SNR shocks in the inferred velocity range (Ghavamian et al. 2007).

\subsection{The Compact Source}

Using the derived age for the remnant, we calculate the projected velocity of \cco\ to be $1400 - 2600{\rm\ km\ s^{-1}}$, 
assuming a displacement of $1.7 d_{4.5}$~pc (1.3~arcmin) from the adopted center of the SNR. 
We generated a lightcurve of \cco\ but found no short-term flux variation within the \chandra\ exposure. We applied barycenter corrections to the photon arrival times and searched for periodicity using the $Z^{2}$-test (Buccheri et al. 1983). No pulsations with period longer than 6.4~s and pulsed fraction larger than 16\% were detected at 99\% confidence. The \chandra\ image in Figure 1a shows no extended emission surrounding the CCO and a generated radial profile (Figure 5) of the source confirms that it is fully consistent with a model PSF. As a note, photon pile-up is negligible with a level of less than 5\%.

\begin{table}
\begin{center}
\caption{\footnotesize{Derived Parameters for the Evolution of \snr}}
\label{tab:dens}
\begin{tabular}{lrr}
\toprule
\toprule

Parameter && Value \\

\midrule
Age, $t$, (years) && $600 - 1200$ \\
\noalign{\smallskip}
Shock Velocity, $v_{b}$, (${\rm km\: s^{-1}}$) && $1500 - 2900$ \\
\noalign{\smallskip}
Proton Temperature, $kT_{p}$, (keV) && $3 - 10$ \\
\noalign{\smallskip}
Speed of CCO, $v_{cco}$, (${\rm km\: s^{-1}}$) && $1400 - 2600$ \\
\bottomrule
\end{tabular}
\end{center}
\end{table}

We extracted the $\sim 4000$ CCO counts from a 2~arcsec-radius aperture and grouped the source spectrum to a minimum of 25
counts per bin. Using the \emph{SHERPA} environment, we fit the spectrum to simple absorbed blackbody (BB) and power-law (PL)
models. Although the PL fit is slightly better than the BB, the large photon index ($\Gamma = 5.5$) suggests a thermal origin
for the emission. Both fits are consistent with those discussed in G08. Since CCOs are often characterized by two-component
models (BB+PL and BB+BB), we investigated those as well. The results are summarized in Table 3. The best-fit absorption
column density is $\sim 4 \times 10^{22}{\rm\ cm^{2}}$ and agrees with that of the SNR. While the CCO spectrum is dominated
by a BB with $kT \approx 0.4$~keV and a small emission radius of $\sim 3$~km, the high energy component is not very well
constrained. Our results suggest that it can be modeled by either an additional BB component with a very high temperature
(0.9~keV), or a PL component. Adding a second component improves the fit
although not at a statistically significant level.  Unfortunately, the current data do not allow us to distinguish between these two scenarios and
a deeper exposure is required. We present the spectrum of \cco\ overlaid with the BB+PL model in Figure 6. 

\begin{table*}
\begin{center}
\caption{\footnotesize{Spectral Fits to \cco}}
\label{tab:dens}
\begin{tabular}{lrrrrr}
\toprule
\toprule

Parameter && BB & PL & BB+PL & BB+BB \\

\midrule
$N_{\rm H}$ ($10^{22}$\,cm$^{-2}$) && $3.4_{-0.1}^{+0.2}$ & $4.4\pm0.1$ & $4.1_{-0.6}^{+1.4}$ & $4.0_{-0.3}^{+0.4}$ \\
\noalign{\smallskip}
$kT_{1}$ (keV) && $0.50\pm0.01$ & \nodata & $0.42_{-0.06}^{+0.05}$ & $0.41\pm0.05$ \\
\noalign{\smallskip}
$R_1$ (km) && $1.59\pm0.02$ & \nodata & $2.59\pm0.06$ & $2.94\pm0.03$ \\
\noalign{\smallskip}
$\Gamma$ && \nodata & $5.5\pm0.3$ & $3_{-5}^{+2}$ & \nodata \\
\noalign{\smallskip}
$kT_2$ (keV) && \nodata & \nodata & \nodata & $0.9_{-0.2}^{+0.8}$ \\
\noalign{\smallskip}
$R_2$ (km) && \nodata & \nodata & \nodata & $0.19\pm0.01$ \\
\noalign{\smallskip}
Flux\tablenotemark{a} (${10^{-13}\,\rm\ ergs\ cm^{-2}\ s^{-1}}$) && $4.4\pm0.1$ & $4.7\pm0.2$ & $4.9\pm0.2$ & $4.7\pm0.1$ \\
\noalign{\smallskip}
$F_1/F_2$\tablenotemark{b} && \nodata & \nodata & 1.5 & 9.8 \\
\noalign{\smallskip}
$\chi^2_\nu$/dof && 1.02/105 & 0.94/105 & 0.86/103 & 0.86/103 \\
\bottomrule
\noalign{\smallskip}
\multicolumn{6}{p{1.1\columnwidth}}{\footnotesize{$^{\text{a}}$The flux is over the energy range $0.5 - 10.0$~keV and has been corrected for foreground absorption}}\\
\multicolumn{6}{p{1.1\columnwidth}}{\footnotesize{$^{\text{b}}$Unabsorbed flux ratio between the first and second spectral components in the $1.0 - 10.0$~keV energy range}}\\
\end{tabular}
\end{center}
\end{table*}

\section{Discussion}

The high resolution \chandra\ data reveal that the emission from \snr\ extends far to the west of the bright ejecta region
and provide a rough constraint on the size of the SNR, with the caveat that the asymmetric morphology makes it difficult to
extract a reliable diameter. G08 estimated the age of \snr\ to be $\sim 900$~years based on the diameter of the ejecta region
alone ($D \approx 2.6 d_{4.5}$~pc) and the shock velocity derived from the electron temperature. However, we see that the
emission from \snr\ extends far beyond the bright ejecta-dominated region and our analysis (using the model treatment of
TM99) shows that the electron temperature is almost certainly an underestimate of the proton temperature. With the diameter,
a post-shock density estimate and an assumed range of values for $M_{ej}$ and $E_{SN}$ as input parameters, we calculate a
shock velocity of $1500 - 2900{\rm\ km\ s^{-1}}$, a proton temperature of $3 - 10$~keV and an age of $600 - 1200$~years. We
see that although the size of the SNR is far larger, the age is consistent with earlier estimates due to the lower ISM
density estimate. 

\begin{figure}[t]
\epsscale{1.18}
\plotone{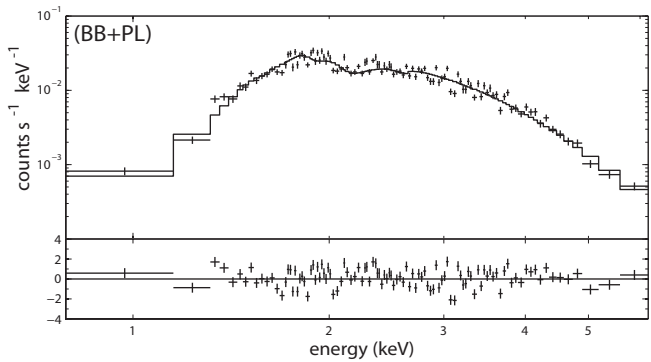}
\caption{
The X-ray spectrum of CCO \cco\ overlaid with a BB+PL model. The fit parameters are shown in Table 3.
}
\end{figure}

The enhanced elemental abundances and high temperature of the bright emission in the remnant's eastern half unambiguously
demonstrate the presence of hot, metal-rich ejecta. By contrast, spectral analysis of the emission around the perimeter of
the ejecta region and throughout the SNR's western half (with the possible exception of region 2 --- see discussion in \S\
3.1) reveals cooler material with solar-type composition, which supports shocked
interstellar/circumstellar medium (ISM/CSM) for its origin. This well-defined
interface between the ejecta and shocked ISM regions further supports G08's suggestion that \snr\ is interacting with a
molecular cloud in the east. If this is indeed the case, the expanding shock wave propagating eastward through the dense gas
would form a strong reverse shock and heat the metal-rich ejecta to X-ray emitting temperatures while the shock front
propagating in the other direction would still be expanding relatively unimpeded into the diffuse ISM. 

Although we cannot conclusively determine whether the 24~${\rm\ \mu m}$ emission is due to dust or line emission, the
similarity between the infrared and X-ray morphologies supports the shocked dust scenario (Seok et al. 2008). In addition,
the lack of significant emission at 70~${\rm\ \mu m}$ and in the IRAC band is consistent with dust emission that peaks between
those bands. The catalog of 24~${\rm\ \mu m}$ MIPSGAL sources by Mizuno et al. (2010) lists only a handful of SNRs, but
spectroscopic observations with \spitzer\ have shown that SNRs interacting with dense molecular clouds produce a number of
shocked molecular and atomic species in the mid-infrared band (Hewitt et al. 2009). Thus the presence of 24~${\rm\ \mu m}$
emission coinciding with the eastern bright X-ray clump in \snr\ is consistent with a molecular cloud interaction. 
As can be seen from Figure 4, the upper filament (region 1b) appears to outline precisely the eastern edge of 
of the region corresponding to the ejecta-dominated emission. By contrast, the
lower filament (region 1a) is 
not visible at 24~${\rm\ \mu m}$. The Mg and Si lines in the spectra of these two regions (Figure 2) 
as well as the spectral fit results (Table 1) suggest that the ionization timescale of the region 1a 
filament is noticeably higher than that of the region 1b filament. Although we cannot determine whether the 
difference is due to a higher density or longer time since shock, the material in region 1a (in 
either case) may have cooled/sputtered more than that in the region 1b filament.
Incidentally, both regions of emission at 24~${\rm\ \mu m}$ (1c and 4) coincide with 
the regions of lowest ionization timescale, as determined from their X-ray
spectra.  This result seems consistent with the scenario that the 24~${\rm\ \mu m}$ emission is due to dust 
heated by the reverse shock propagating back into the SNR and the difference in brightness between the various regions
could be attributed to dust cooling or sputtering.

If the explosion occurred approximately at the center of the X-ray emission (RA. 17:21:01.019, DEC. -37:26:49.61), the
derived age puts the projected velocity of \cco\ at $1400 - 2600{\rm\ km\ s}^{-1}$, assuming it was formed in the same
explosion. The lower limit of this velocity, although high, falls within the $1100 - 1600{\rm\ km\ s^{-1}}$ measured for the
CCO in Puppis A (Hui \& Becker 2006: Winkler \& Petre 2007). The upper limit is significantly higher than that measured for
any CCO. Although the spectrum of \cco\ can be fitted with a BB+BB or a BB+PL model, the former requires a BB component of
unusually high temperature. On the other hand, a BB+PL model is not uncommon when comparing to the X-ray spectra of other
CCOs. The best-fit BB+PL parameters in Table 3 fall within the typical range of others (see Table 2 in Pavlov et al. 2004).
Indeed, \cco\ could be a close cousin of the CCOs inside SNRs Vela Jr (G266.1-1.2) and Puppis A. These objects have an age of
$1 - 3$~kyr and a blackbody temperature of $\sim0.4$~keV. We also note that for the CCO inside Puppis A, \emph{XMM-Newton}
observations indicate an additional hard spectral component that can be fitted with either a PL of $\Gamma = 2.0 - 2.7$ or a
hard BB with $kT = 0.5 - 1.1$~keV (Becker \& Aschanbach 2002), very similar to our case.

In addition to the prominent difference in flux between the bright eastern emission and the surrounding regions, there is an
intriguing contrast in brightness between the ambient material in the far west and the bright regions of shocked CSM/ISM
(regions 3 and 4 in Figure 1a). We estimate the brightness ratio of these regions to be $\sim 3$, implying a density ratio of
$\sim 1.7$. Thus, if more of the remnant lies unseen further to the west, it would need to be only slightly less dense to
emit below the background. If this is indeed the case, it would change all the derived
SNR parameters and imply that the SNR may have already entered the ST phase. The larger radius would increase the SNR's
age, lower the shock velocity and lower the proton temperature. In addition, it would imply that \cco\
may be closer to the explosion center than evident from the visible emission, which would place its speed at a
lower, more typical value. On the other hand, if \snr\ is indeed interacting with a molecular cloud along its eastern edge,
the apparent center of the emission is likely a poor estimate of the explosion location. In this case, the supernova center
would be closer to the eastern edge, which would tend to increase the velocity of the CCO. 

\section{Conclusion}

In this paper, we have investigated the dynamical properties of \snr\ and its candidate neutron star \cco. By doing spectral
modeling on the high resolution X-ray data, we have found that the region of brightest emission is dominated by hot,
metal-rich ejecta while the diffuse material throughout the surrounding regions is mostly cooler swept-up CSM/ISM with solar
abundances. These results further support the conclusion of Gaensler et al.
(2008) that \snr\ is interacting with a dense molecular cloud in the
east. The X-ray imaging has resolved new morphological features and revealed that the SNR is far more extended than apparent
from the ejecta region alone. We used the numerical model of Truelove and McKee
(1999) to show that \snr\ is likely in the free expansion phase
and derive values for its age, shock velocity and proton temperature. We examined the relationship between the inferred shock
velocity and the derived electron/proton temperature ratio for \snr\ and found it to be entirely consistent with that of
other SNRs. The derived age puts the speed of \cco\ at an unusually high value although its spectral characteristics are
found to be consistent with CCOs in other remnants. We presented an archival 24~${\rm\ \mu m}$ image of \snr\ and found that
the infrared and X-ray morphologies are well-correlated, offering additional support for a molecular cloud interaction. 

\acknowledgments

The authors would like to thank Tea Temim and Daniel Castro for their input and
several helpful discussions. I.L. acknowledges support from \emph{Chandra} grant
GO9-0059X. P.O.S. acknowledges partial support from NASA contract NAS8-03060.
J.D.G is supported by an NSF Astronomy and Astrophysics Postdoctoral
Fellowship under award AST-0702957. B.M.G. acknowledges the support of the Australian Research Council through grant FF0561298.


\end{document}